# Custom Deep Neural Network for 3D Covid Chest CT-scan Classification

Quoc-Huy Trinh
Ho Chi Minh University of Science
VNU-HCM, Ho Chi Minh city, Vietnam
trnhquchuy@yahoo.com.vn

Minh-Van Nguyen
Ho Chi Minh University of Science
VNU-HCM, Ho Chi Minh city, Vietnam
20120709@student.hcmus.edu.vn

## Abstract

*3D CT-scan base on chest is one of the controversial topisc of the researcher nowadays. There are many tasks to diagnose the disease through CT-scan images, include Covid19[6]. In this paper, we propose a method that custom and combine Deep Neural Network to classify the series of 3D CT-scans chest images. In our methods, we experiment with 2 backbones is DenseNet 121 and ResNet 101. In this proposal, we separate the experiment into 2 tasks, one is for 2 backbones combination of ResNet and DenseNet, one is for DenseNet backbones combination.*

## 1. Introduction

The goal of the 3D-CT scans images classification task is to evaluate various methods to classify the 3D-CT scans images correctly and efficiency[4]. In this paper, we propose a method that ensemble Deep Neural Network backbones to classify 3D-CT scans images. In this experiment, we use backbones DenseNet 121, ResNet 101 for evaluate our method on the test dataset.

### 1.1. Dataset

The Dataset we use is from MIA-COVID 19 dataset . The dataset contain which Covid 3D-CT Scan images from patients that have COVID 19 and the patient that do not have COVID 19. The dataset is splitted into folders, each folder is the series of images when doing CT-Scan.[3]

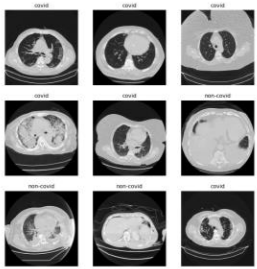

Figure 1. Sample of Dataset

### 1.2. Evaluation methods

To evaluate the method, the challenge use Macro F1Score with the following formula [7]:

$$\frac{1}{n} * \sum_{i=0}^{n} F1 - scores_i$$

Where:

n: number of classes/labels

i : class/label

## 2. Method

In spite of using RNN and LSTM to find the feature of 3D- CT Scan series, we propose a method that extract all features of all images in all series[5]. This will efficiently reduce the time of training. In the testing phase, we propose to predict all images in the series and calculate the average score of the serie to choose the label of that serie.

### 2.1. Densely Connected Convolutional Network

Recent work has shown that convolutional networks can be substantially deeper, more accurate, and efficient to train if they contain shorter connections between layers close to the input and those close to the output.DenseNet which connects each layer to every other layer in a feed-forward fashion. Whereas traditional convolutional networks with L layers have L connections - one between each layer and its subsequent layer - our network has L(L+1)/2 direct connections. For each layer, the feature-maps of all preceding layers are used as inputs, and its own feature-maps are used as inputs into all subsequent layers. DenseNets have several compelling advantages: they alleviate the vanishinggradient problem, strengthen feature propagation, encourage feature reuse, and substantially reduce the number of parameters.[2]

## 2.2. Deep Residual Network (ResNet)

Deeper neural networks are more difficult to train. Deep Residual Network is created to ease the training of networks that are substantially deeper than those used previously. ResNet can explicitly reformulate the layers as learning residual functions with reference to the layer inputs, instead of learning unreferenced functions. With this network, it is easier to optimize, and can gain accuracy from considerably increased depth. On the ImageNet dataset the evaluation of residual nets with a depth of up to 152 layers—8x deeper than VGG nets but still having lower complexity. An ensemble of these residual nets achieves 3.57 error on the ImageNet test set. This result won the 1st place on the ILSVRC 2015 classification task. We also present analysis on CIFAR-10 with 100 and 1000 layers.[1]

## 2.3. Res-Dense Net

In our architecture, we propose to use ResNet 101 and DenseNet 121 backbones for the first layers. We will have two pipes: ResNet 101 and DenseNet. The output of ResNet 101 will be extracted, by a Conv2D, to have the same shape as DenseNet 121 output. After feature extraction, they are added to create the feature map before coming to global Average Pooling layers for being classified.

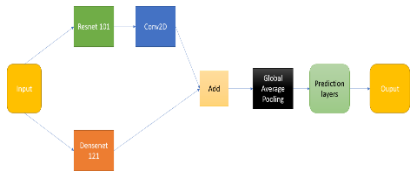

Figure 2. Network Architecture

After 2 Extraction layers, the feature map of the images is briefer, ease for classification process.

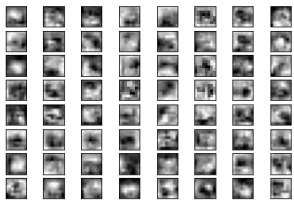

Figure 3. Feature extraction of Res-Dense Net

## 2.4. Data preprocessing

After loading data, we resize all the images to the size (256,256), then we split the dataset into the training set and validation set in the ratio of 0.75:0.25. After resizing and splitting the validation set, we rescale the data pixel down to be in the range [-1,1] by divide by 127.5. Then we use the application of ResNet to preprocess input.

## 2.5. Data Augmentation

To reduce the Overfitting problem, we use augmentation to generate the data randomly by random flip images and random rotation with an index of 0.2.

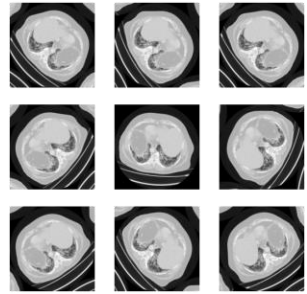

Figure 4. Data Augmentation result

## 2.6. Training

Our models are initialized with pre-trained weight from Tensorflow Imagenet. We use a batch size of 32 for training data with an image's size of (256,256). We use RMSprop with a learning rate is 0.0001 for optimizer and evaluate the training process by accuracy and F1-score. For the loss function, we use Sparse Categorical Cross-entropy. We train the model with 20 epochs and get the checkpoint that has the highest validation loss. Firstly, we freeze all the complicated layers of DenseNet and ResNet. Then, we start to train for the first time, then we freeze 100 layers before and we start the continuous training process.

## 3. Conclusion

We demonstrated the proposal of using Res-Dense Net with Fine-tuning technique to classify endoscopic images. The result of our research is positive for F1 score. However, there are some drawbacks that we have to do to improve the performance of the model, such as pre-processing data, reduce noise, change the size of the image to train. Furthermore, we can apply ResNet101 V2 or DenseNet 169 backbone, or we can combine with LSTM or RNN modules to have better feature extraction and better performance of the model.